# Discovery of two families of VSb-based compounds with V-kagome lattice

Yuxin Yang,[1,2,†] Wenhui Fan,[1,2,†] Qinghua Zhang,[1,†] Zhaoxu Chen,[1,2] Xu Chen,[1,2] Tianping Ying,[1,3]* Xianxin Wu,[4] Xiaofan Yang,[5] Fanqi Meng,[6] Gang Li,[1,7] Shiyan Li,[5] Tian Qian,[1,7] Andreas P. Schnyder,[4] Jian-gang Guo,[1,7]* Xiaolong Chen[1,7]*

[1]Beijing National Laboratory for Condensed Matter Physics, Institute of Physics, Chinese Academy of Sciences, Beijing 100190, China

[2]School of Physical Sciences, University of Chinese Academy of Sciences, Beijing 100049, China

[3]Materials Research Center for Element Strategy, Tokyo Institute of Technology, 4259 Nagatsuta, Midori, Yokohama 226-8503, Japan

[4]Max-Planck-Institut für Festkörperforschung, Heisenbergstrasse 1, D-70569 Stuttgart, Germany

[5] State Key Laboratory of Surface Physics, Department of Physics, Fudan University, Shanghai 200438, China.

[6] School of materials, Tsinghua University, Beijing 100084, China

[7]Songshan Lake Materials Laboratory, Dongguan 523808, China

(Dated: Nov. 18th, 2021)

**Abstract**

We report the structure and physical properties of two newly-discovered compounds $AV_8Sb_{12}$ and $AV_6Sb_6$ (A = Cs, Rb), which have $C_2$ (space group: Cmmm) and $C_3$ (space group: R-3m) symmetry, respectively. The basic V-kagome unit is present in both compounds, but stacking differently. A $V_2Sb_2$ layer is sandwiched between two $V_3Sb_5$ layers in $AV_8Sb_{12}$, altering the V-kagome lattice and lowering the symmetry of kagome layer from hexagonal to orthorhombic. In $AV_6Sb_6$, the building block is a more complex slab made up of two half-$V_3Sb_5$ layers that are intercalated by Cs cations along the c-axis. Transport property measurements demonstrate that both compounds are nonmagnetic metals, with carrier concentrations at around $10^{21}cm^{-3}$. No superconductivity has been observed in $CsV_8Sb_{12}$ above 0.3 K under in-situ pressure up to 46 GPa. In contrast to $CsV_3Sb_5$, theoretical calculations and angle-resolved photoemission spectroscopy (ARPES) reveal a quasi-two-dimensional electronic structure in $CsV_8Sb_{12}$ with $C_2$ symmetry and no van Hove singularities near the Fermi level. Our findings will stimulate more research into V-based kagome quantum materials.

**Introduction**

The kagome net, a corner-sharing tiling pattern of triangular plaquettes, is a standard model for understanding nontrivial topology, frustrated magnetism, and correlated phenomena. The frustrated lattice geometry endows a wealth of unique physical features to the lately intensely researched kagome metals, including flat bands, van Hove singularity, and topological band crossover. Pure kagome layers, on the other hand, cannot stand alone and are usually intergrown with other building blocks. The stacking-induced interaction and charge donation bring rich varieties to the band structures inherited from the kagome lattice. Prominent examples are noncollinear antiferromagnetism in $Mn_3Sn$ [1], Dirac fermionic in ferromagnetism in $Fe_3Sn_2$ [2], Chern-gapped Dirac fermion in ferromagnetic $TbMn_6Sn_6$ [3], and Weyl fermions in the ferromagnet $Co_3Sn_2S_2$ [4].

The seminal discovery of $AV_3Sb_5$ [5,6] has inspired a new surge of interest. As seen in Fig. 1, the vanadium sublattice forms a perfect kagome net that is interwoven with Sb atoms both within and outside of the plane to build a $V_3Sb_5$ monolayer. Despite the lack of magnetization, $AV_3Sb_5$ is discovered to be $Z_2$ topological materials with alternative ground states such as the charge density wave (CDW) ordering and superconductivity. This CDW ordering is later proved not driven by strong electron-phonon coupling [7–14] and competes with superconductivity [15]. Subsequent studies further discovered exciting discoveries including the reentrance of superconductivity under high pressure [16–19], the enormous anomalous Hall effect [20,21], pair density wave [11], and evidence of plausible unconventional superconductivity [11,22]. Changing the stacking sequence of the kagome layer with different building components and seeing the consequences on physical characteristics is both exciting and practically doable, given the very weak interlayer interactions. However, only partial removal A element has been realized through selective oxidation of thin flakes, resulting in a shift of van Hove singularity [23,24], enhanced $T_c$ [15,25–28], and A-vacancy ordering [29]. The construction of novel vanadium-based kagome structures by altering the stacking sequence has not been realized so far.

Here, we investigate the possibility of constructing novel A-V-Sb compounds by rearranging the constituent layers. As a result, the $AV_8Sb_{12}$ and $AV_6Sb_6$ (A = Rb, Cs) families of compounds are found. The initial $C_6$ symmetry is decreased to $C_2$ and $C_3$ in $AV_8Sb_{12}$ and $AV_6Sb_6$, respectively, by inserting alternate building blocks or changing the stacking sequences. Their band structures and electric conductivity are substantially altered when the symmetry is broken. Theoretical calculations and ARPES measurements reveal previously unseen features compared to that of $AV_3Sb_5$.

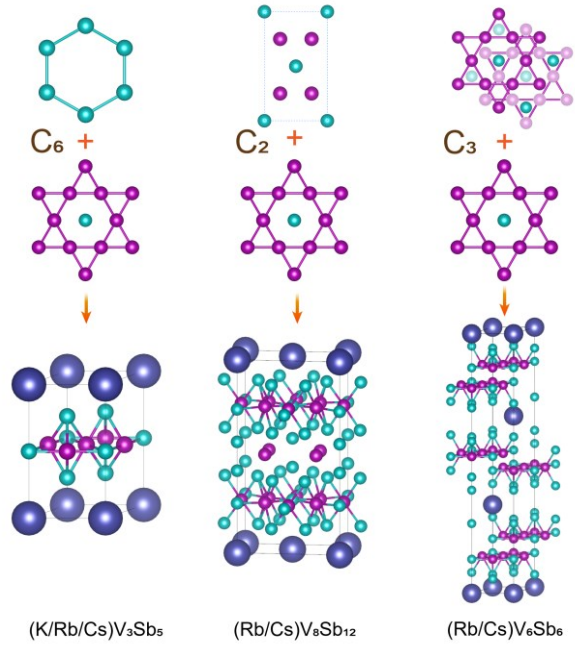

FIG. 1. Crystal structure of $AV_3Sb_5$, $AV_8Sb_{12}$, and $AV_6Sb_6$ (A=K, Rb, Cs). Blue, violet and green balls are A, V, and Sb atoms, respectively.

**Experimental**

**Single crystal growth**: Single crystals of $AV_8Sb_{12}$ and $AV_6Sb_6$ (A = Cs, Rb) were grown via the self-flux method. $AV_8Sb_{12}$ ($AV_6Sb_6$) single crystals were grown by mixing Cs (99.98%), V (powder, 99.99%) and Sb (ingot, 99.999%) with the molar ratio of 1.2:6:15 (1.2:6:18), loaded into an alumina container and then sealed into a silica tube in vacuum. The excess flux was removed by centrifuging at 1173 K. For Cs (99.98%), V (powder, 99.99%), and Sb (ingot, 99.999%) were mixed with the molar ratio, loaded into an alumina container and then sealed into a silica tube in vacuum. The mixture was subsequently heated to 1373 K and kept for 24 h, then cooled to 1273 K in 72 h. The excess flux was removed by centrifuging at 1273 K.

**Characterization:** The diffraction was performed on Panalytical X'pert diffractometer with a Cu $K\alpha$ anode (1.5481 Å). The composition and structure of the sample were determined by the combination of EDS and scanning transmission electron microscope (STEM). The atomic arrangement of the two-phase was observed by spherical aberration-corrected ARM200F (JEOL, Tokyo, Japan) STEM operated at 200 kV with a convergence angle of 25 mrad and collected angle from 70 to 250 mrad. The high angle annular dark field STEM (HAADF-STEM) image was collected with a dwell time of 10 μs each pixel. The transport measurement under ambient and high pressure was performed on Quantum Design Physical Property Measurement System (PPMS). A Keithley 2182A, a Keithley 6221, and a Keithley 2400 were used to measure the transport properties under external magnetic fields. Magnetization measurements were performed using a Quantum Design Magnetic Properties Measurement System (MPMS3). The high-pressure resistivity of $CsV_8Sb_{12}$ samples was measured by a diamond anvil cell (DAC) range from 2K to 400K with the van der Pauw method. The resistance experiments were performed using Be–Cu cells. The cubic boron nitride (cBN) powders (200 and 300 nm in diameter) were employed as the medium to transfer pressure. The pressure was calibrated using the ruby fluorescence method at room temperature, whether before and after the measurement.

**Angle-resolved photoemission spectroscopy.** All the ARPES data shown are recorded at the "Dreamline" beamline of the Shanghai Synchrotron Radiation Facility. The energy and angular resolutions are set to 15 to 25 meV and 0.2°, respectively. All the samples for ARPES measurements are mounted in a BIP Argon (>99.9999%)–filled glove box, cleaved in situ, and measured at 25 K in a vacuum better than $5 \times 10^{-11}$ torr.

**Computational:** Our DFT calculations employ the projector augmented wave (PAW) method [30] encoded in Vienna ab initio simulation package (VASP) [31], and both the local density approximation (LDA) and generalized-gradient approximation (GGA) [32] for the exchange-correlation functional are used. Throughout this work, the cutoff energy of 500 eV is taken for expanding the wave functions into a plane-wave basis. In the calculation, the Brillouin zone is sampled in the k space within the Monkhorst-Pack scheme [33]. The number of these k points is 9x9x5 for the primitive cell. We relax the lattice constants and internal atomic positions, where the plane wave cutoff energy is 600 eV. Forces are minimized to less than 0.01 eV/Å in the relaxation. The obtained lattice constants of $CsV_8Sb_{12}$ are $a$=9.515 Å, $b$=5.455 Å and $c$=18.25 Å, which are in good agreement with experimental values $a$=9.516 Å, $b$=5.451 Å, and $c$=18.128 Å.

**Results and discussions**

The crystal structures of $AV_8Sb_{12}$ and $AV_6Sb_6$ are obtained using a combination of x-ray diffraction and scanning transmission electron microscope (STEM), as shown in Fig. 1, together with the structure of $AV_3Sb_5$ for comparison. The three A-V-Sb compounds all have a V-based kagome lattice with a centered Sb atom as a common building block. Each V-based kagome layer is sandwiched by two sets of Sb honeycomb lattices in the $AV_3Sb_5$, which has a space group of *P*6/mmm with $C_6$ symmetry. The crystal structure is rather simple, with the lattice constants of $a$=5.4922 Å and $c$=9.8887 Å. $AV_8Sb_{12}$ and $AV_6Sb_6$, on the other hand, reveal richer crystal structures due to the intercalation of low symmetric structural units. In $AV_8Sb_{12}$, an orthorhombic $V_2Sb_2$ layer is sandwiched between two $V_3Sb_5$ units, forming a $V_8V_{12}$ unit. The space group is Cmmm with lattice constants of $a$=5.4510 Å, $b$=9.5164 Å, and $c$=18.1282 Å. The $a$-axis is approximately √3 times that of $AV_3Sb_5$. As for $AV_6Sb_6$, the basic unit can be viewed as a two-$V_3Sb_5$-connected slab by deleting the middle two Sb layers. Each half-$V_3Sb_5$ unit is displaced by (1/2, 1/2, 0), and the Cs atoms run down the $c$-axis to separate the slabs. The space group is *R*-3m with $a$=5.3172 Å and $c$= 34.0741 Å, in which the $c$-axis is almost 3 times that of $AV_3Sb_5$. $AV_8Sb_{12}$ and $AV_6Sb_6$ can be viewed as derivative phases of the $AV_3Sb_5$ phase. Since the V-based kagome lattice is mostly intact, exotic properties like superconductivity and charge order are highly expected.

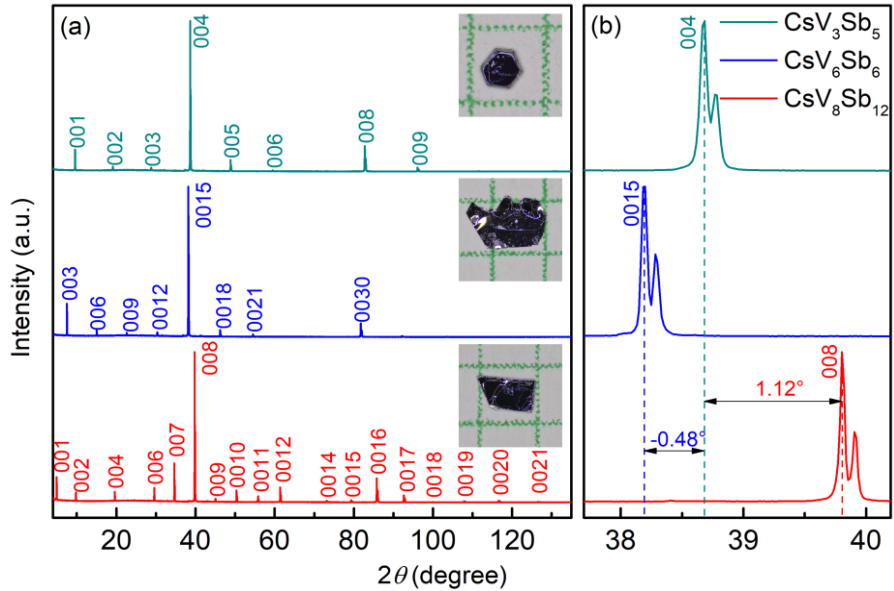

Fig. 2. (a) XRD patterns of CsV$_3$Sb$_5$, CsV$_8$Sb$_{12}$ and CsV$_6$Sb$_6$. Right panel shows the zoomed-in peaks (004), (0015) and (008) from ~38° to ~40°.

We have grown the single crystals of three A-V-Sb compounds (insets of Fig. 2a). All three samples have a layered structure with metallic luster and a shining surface. From the XRD patterns, we can index the lattice constant *c* based on the 00*l* peak. The calculated *c* values in the three compounds agree well with the proposed lattice constants after structural relaxation. We further perform atomic resolved STEM imaging and EELS spectra mapping along the two main zone axis: [100] and [110]. The atomic arrangements in each layer can be identified clearly, as shown in Fig. 3a and Fig. 3b. The elemental mapping along the [100] projected also verified this double kagome lattice at an atomic scale. The false-color image of elemental mapping of AV$_8$Sb$_{12}$ (AV$_6$Sb$_6$) is shown in Fig. 3(c,f), from which we can see the atomic arrangements match the proposed crystal structure of Fig. 1(b).

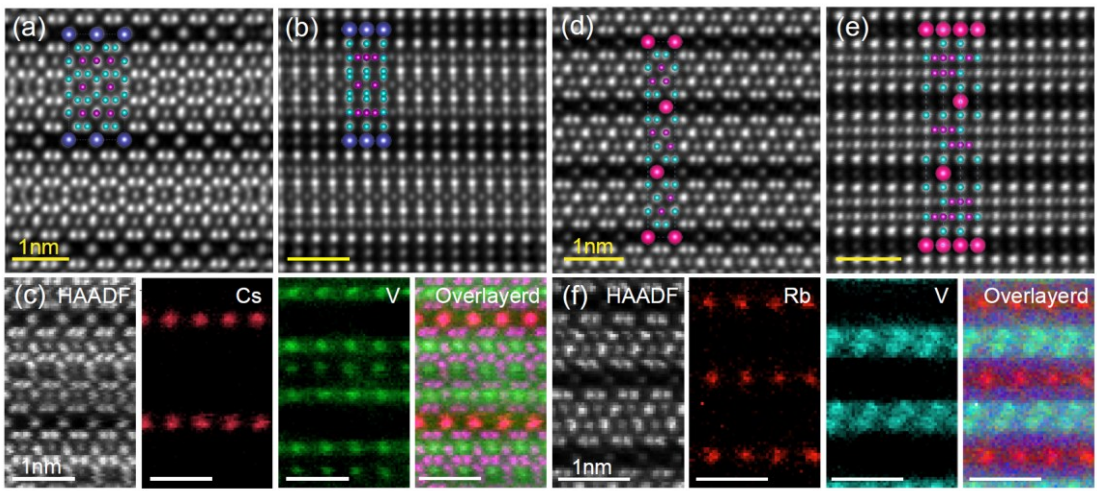

FIG. 3. Atomic structures of $CsV_8Sb_{12}$ and $RbV_6Sb_6$ are characterized by STEM. The HAADF image along the [100] (a) and [110] (b) projection, where the structure model is overlayed to show the atomic arrangement. (c)The elemental mapping of $CsV_8Sb_{12}$ is based on the EELS spectra. The overlayed image is composed of Red Cs, green V, and purple HAADF contrast. The HAADF image along the [100] (d) and [110] (e) projection of $RbV_6Sb_6$, where the structure model of $RbV_6Sb_6$ is overlayed to show the atomic arrangement. (f)The elemental mapping of $RbV_6Sb_6$ is based on the EELS spectra. The overlayed image is composed of Red Rb, green V, and purple HAADF contrast.

We use $CsV_8Sb_{12}$ and $RbV_6Sb_6$ to represent the general behavior of the two newly discovered families. Both samples exhibit a metallic behavior from 300 to 2 K without any anomalies observed (Fig. 4a). We have also measured $CsV_8Sb_{12}$ and $CsV_6Sb_6$ down to 0.3 K and 100 mK, respectively. No superconductivity can be observed. The residual-resistance ratios (RRR) are generally low, with the highest value of 2.8 for $CsV_8Sb_{12}$. Figure 4b shows the magnetization of $CsV_8Sb_{12}$ and $RbV_6Sb_6$. Both samples exhibit Pauli magnetism down to 100 K. The upturn at low temperatures may originate from magnetic impurities. Both in and perpendicular to the ab plane, the overall resistivity of $CsV_8Sb_{12}$ is nearly four times higher than that of $RbV_6Sb_6$.

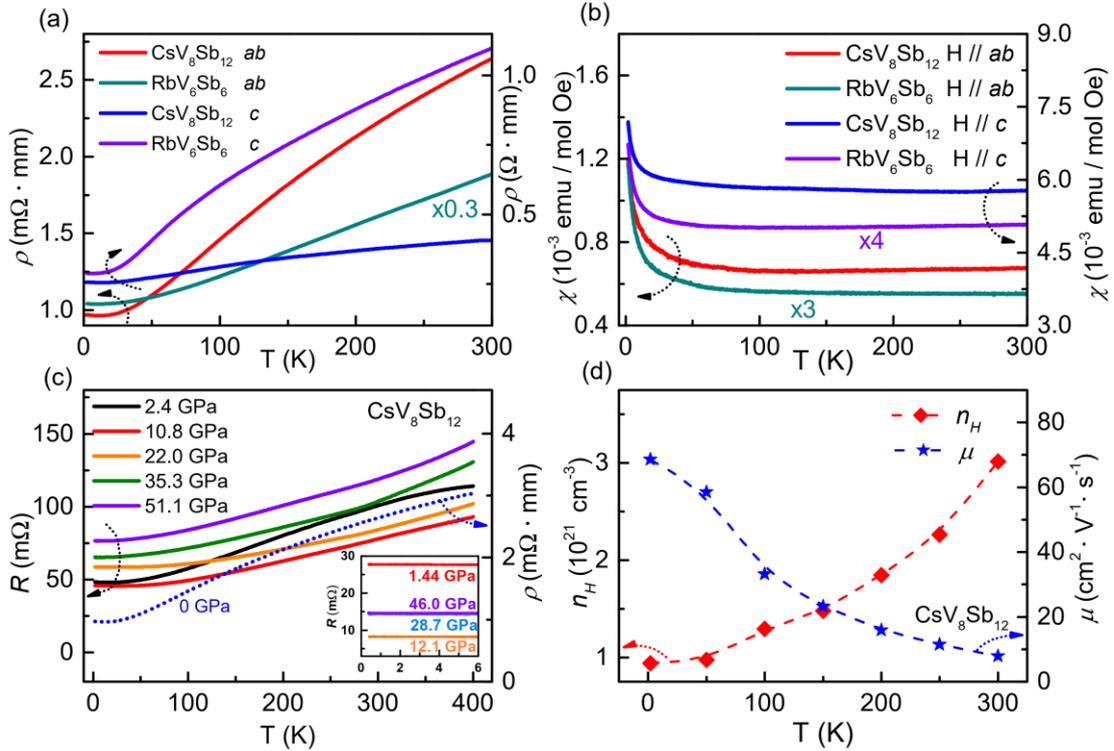

Fig. 4. Electrical transport and magnetization of $CsV_8Sb_{12}$ and $RbV_6Sb_6$. (a, b) Temperature-dependent resistivity and magnetization of $CsV_8Sb_{12}$ and $RbV_6Sb_6$ within the *ab* plane and along the *c*-axis. (c) Temperature-dependent resistance of $CsV_8Sb_{12}$ under different external pressure. The measured temperature ranges from 2-400 K. We superimposed the resistivity curve at ambient pressure as the dotted line. Inset is the high-pressure measurements down to 0.3 K. (d) Carrier concentration and mobility of $CsV_8Sb_{12}$ against temperature.

The recently discovered reentrance of superconductivity in CsV$_3$Sb$_5$ provoke us to further investigate the influence of external pressure on the electric transport properties in AV$_8$Sb$_{12}$ and AV$_6$Sb$_6$. As shown in Fig. 4c, the resistance initially decreases from 2.4 to 4.8 GPa. With further increasing the applied pressure, the resistance gradually increases to a high value. We note that the shape of the RT curve at 2.4 GPa resembles that of resistivity measured at ambient pressure. A prominent feature is the RT curves change from a convex to concave over 10.8 GPa, indicating a dramatic change of the electrical properties. However, our high-pressure and low-temperature measurements do not show any sign of superconductivity down to 0.3 K (inset of Fig. 4c). Despite the relatively low *RRR* value found in CsV$_8$Sb$_{12}$, the extracted carrier concentration and electron mobility are comparable to that of CsV$_3$Sb$_5$. Thus, the absence of superconductivity may be closely related to the insertion of the *C*$_2$-V$_2$Sb$_2$ layer, which may distort the kagome lattice and alter the pairing mechanism.

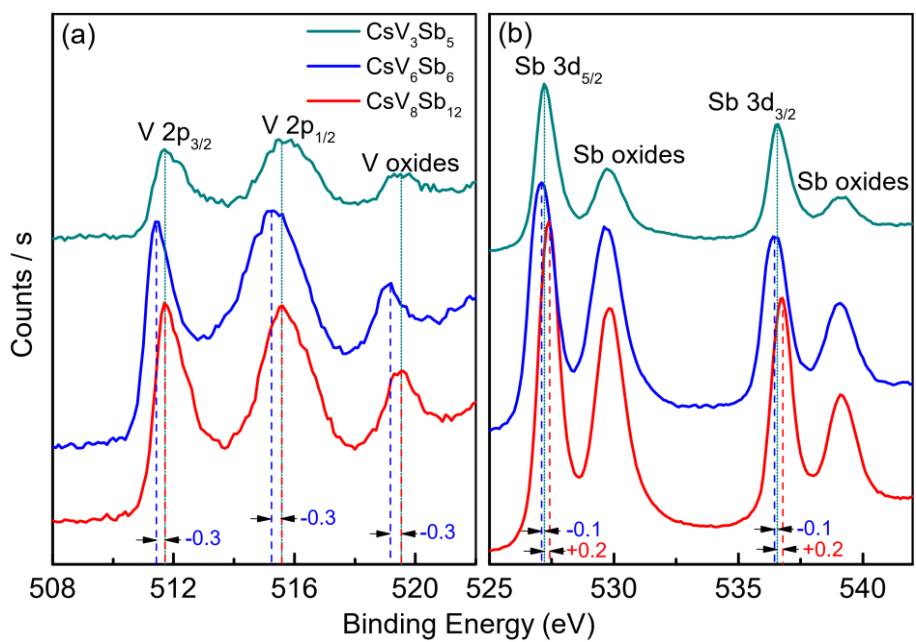

Fig. 5. X-ray photoelectron spectroscopy (XPS) of V-2*p* (a) and Sb-3*d* (b) for CsV$_3$Sb$_5$, CsV$_6$Sb$_6$, and CsV$_8$Sb$_{12}$.

The variation of stacking sequence should be directly reflected in the valence state of the investigated elements. To substantiate the influence of the acquired double-layered kagome compounds, we measured the XPS of CsV$_3$Sb$_5$, CsV$_6$Sb$_6$, and CsV$_8$Sb$_{12}$, as shown in Fig. 5. The peak positions of Cs are identical for all three compounds at 723.8 eV, indicating the complete loss of out shell electrons in all three compounds. An interesting discovery is the opposite evolution of V's and Sb's valence state in CsV$_6$Sb$_6$ and CsV$_8$Sb$_{12}$. In CsV$_6$Sb$_6$, the valence state of V-2*p* shifts to lower binding energy, and the peak of Sb-3d remain unchanged. Meanwhile, the valence state of V remains intact in CsV$_8$Sb$_{12}$, which is accompanied by the noticeable peak shift of the Sb-3d towards higher binding energy. This dramatic difference lies in the fundamental structural difference in their crystal structures. From Fig. 1, the $V_3Sb_5$ kagome layer is identical in both $CsV_3Sb_5$ and $CsV_8Sb_{12}$, while the building block in $CsV_6Sb_6$ evolves into $V_3Sb_3$ with the loss of two Sb atoms on one side. This modification of the kagome layer directly alters the valence state of the V. On the contrary, the inserted building block of $V_2Sb_2$ in $CsV_8Sb_{12}$ connects with the inside V indirectly. Valence modulation of vanadium in the A-V-Sb system through the judicious design of the stacking sequence is realized.

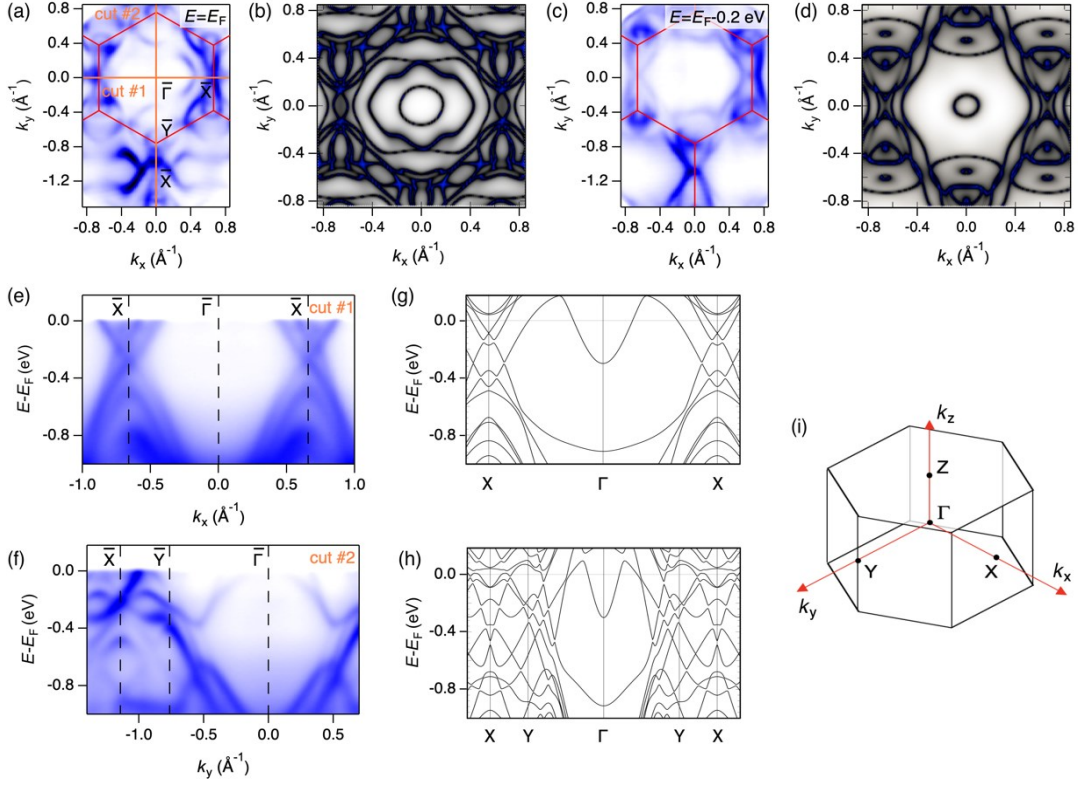

Figure 6. (a,c) ARPES intensity plots at fermi surface and constant energy contours at $E_\text{F} - 0.2$ eV recorded on the (010) surface with $h\nu = 84$ eV. (b,d) DFT calculation at $E_\text{F}$ and $E_\text{F} - 0.2$ eV. (e,f) ARPES intensity plots showing band dispersions along $\bar{\Gamma} - \bar{X}$, $\bar{\Gamma} - \bar{Y}$. (g,h) Calculated band structure along the Γ-X, Γ-Y. (i) 3D bulk Brillouin zone of $CsV_8Sb_{12}$.

To determine the band structure of $CsV_8Sb_{12}$, we have carried out systematical ARPES measurements on the (001) cleavage surface of $CsV_8Sb_{12}$ single crystal. We summarized constant energy contours and band dispersion in Fig. 6. In-plane Fermi surface (FS) measured with $h\nu = 84$ eV in Fig. 6a clearly shows the $C_2$ symmetry feature which got verified in the calculated FS (Fig. 6b), and this testifies that the insert $V_2Sb_2$ layer has strong modulation on the overall band structure. Constant energy contour (Fig. 6c) at 0.2 eV below Fermi level has a good agreement with DFT calculation in Fig. 6d. We also measured the band dispersion along $\bar{\Gamma} - \bar{X}$ and $\bar{\Gamma} - \bar{Y}$, as shown in Fig. 6(e) and 6(f), respectively, whose momentum locations are indicated in Fig. 6(a). These results match our DFT calculations (Fig. 6g and 6h). The absence of the electron bands near the gamma point may be caused by the matrix element effect or possible hole-doping of surface Cs loss.

Due to the $V_2Sb_2$ layer between kagome layers, the point group of $CsV_8Sb_{12}$ is $D_{2h}$, where only a two-fold rotational symmetry persists, in contrast to $CsV_3Sb_5$. Near the Fermi level, the states are dominantly attributed to V 3$d$ orbitals, from both the kagome V layer and the intercalated $V_2Sb_2$ layer. The electron-like band around $\Gamma$ point is attributed to $p_z$ orbitals of Sb atoms in the kagome layer, similar to $CsV_3Sb_5$. The two hole-bands around $\Gamma$ point are mainly contributed by the V 3$d$ orbitals in the $V_2Sb_2$ layer. The band at -0.9 eV around $\Gamma$ point is mainly attributed to V $d$ orbitals from both kagome and intercalated layers.

**Conclusion**

In summary, we have discovered two new families of VSb-based layered compounds, which possess a basic V-kagome lattice similar to that of $AV_3Sb_5$. The intercalation of the $V_2Sb_2$ layer and reorganization of the half-$V_3Sb_5$ layer leads to a lower symmetry of $C_2$ (Cmmm) and $C_3$ (R-3m) compare to $C_6$ of $AV_3Sb_5$. The complex of V-V and V-Sb bonding in three-dimensional space increases the diversity of the VSb-based phase. Exertion of thin-flake engineering or chemical substitution may squeeze out more exotic properties like superconductivity and CDW. Furthermore, the frustration of magnetism and non-trivial topological phenomena are also highly expected in more V-kagome-based compounds.

Note: During the preparation of the manuscript, we became aware of two independent works on $CsV_6Sb_6$ [arXiv:2110.09782] and $CsV_8Sb_{12}$ [arXiv:2110.11452].

Acknowledgments
This work is financially supported by the National Key Research and Development Program of China (No. 2018YFE0202601 and 2017YFA0304700), the National Natural Science Foundation of China (No. 51922105, 11804184, 11974208, and 51772322), Beijing Natural Science Foundation (Grant No. Z200005), and the Shandong Provincial Natural Science Foundation (ZR2020YQ05, ZR2019MA054, 2019KJJ020).

† These authors contribute equally.
T.Y.: ying@iphy.ac.cn
J.G.: jgguo@iphy.ac.cn
X.C.: chenx29@iphy.ac.cn